\documentstyle[eqsecnum,preprint,aps,psfig]{revtex}
\tightenlines
\def\beq{\begin{equation}}
\def\eeq{\end{equation}}
\def\bea{\begin{eqnarray}}
\def\eea{\end{eqnarray}}
\def\nnu{\nonumber}
\def\tst{\textstyle}

\def\al{\alpha}
\def\be{\beta}
\def\gam{\gamma}

\def\eps{\epsilon}
\def\tta{\theta}

\def\om{\omega}

\def\Dta{\Delta}

\def\ptl{\partial}
\def\hf{{1\over2}}
\def\tshf{\tst\hf}

\def\lp{\left(}
\def\rp{\right)}

\def\ham{{\cal H}}
\def\ket#1{|#1\rangle}

\def\tran#1#2{\langle#1|#2\rangle}

\def\xcl{x_{\rm cl}}

\def\bS{{\bf S}}

\def\e0k{E_0^{(k)}}

\begin{document}
\draft

\title{Tunnel splittings for one dimensional potential wells revisited}

\author{Anupam Garg}
\address{Department of Physics and Astronomy, Northwestern University,
Evanston, Illinois 60208}

\date{\today}

\maketitle

\begin{abstract}
The WKB and instanton answers for the tunnel splitting of the ground
state in a symmetric double well potential are both reduced to an
expression involving only the functionals of the potential, without
the need for solving any auxilliary problems. This formula is applied
to simple model problems. The prefactor for the
splitting in the text book by Landau and Lifshitz is amended so as
to apply to the ground and low lying excited states.
\end{abstract}

\widetext

\section{Introduction}
\label{Intro}

The purpose of this article is to reexamine some formal aspects of the problem of
calculating the quantum mechanical tunnel splittings in a smooth,
symmetric, one-dimensional double well potential, such as that
in Fig. 1.  Readers may justifiably wonder if anything new remains
to be said on so mature a subject, and we assure them that,
by and large, there isn't. The physical phenomenon is certainly
very well understood, as are the basic mathematical ideas behind
the calculations. We find, nevertheless, that especially in the case
of the ground state splitting, there is some confusion regarding the
correct WKB answer for this splitting, which is often taken in a form
that is less accurate than it needs to be.  In particular, we
note that the formula for the splitting in the masterly text book by Landau and
Lifshitz \cite{ll} leads to a prefactor which is incorrect for the ground
state \cite{fn1}. 
A second problem is that the form in which this result
is usually presented is poorly suited
to calculation of the ground state splitting. In applying it to model
problems such as the quartic double well potential, e.g., the unwary
student is
unnecessarlily led into asymptotic expansions of elliptic integrals,
which must then be taken from standard tables of formulas.
Because of this confusion, there is also confusion regarding the
equivalence of the WKB \cite{ll} and the instanton \cite{jl,sc}
methods for calculating tunnel splittings.

Like many other problems in physics of a similar nature, the above
discrepancies and their resolution are at the same time ``well-known"
and not known at all!  They are well-known to some experts. Thus, the
correct answers for the splitting are implicit
throughout early WKB studies of the anharmonic oscillator \cite{bw,gp}, 
and can be ferreted out with some work. They are also clearly known to
many field theorists \cite{sc,bw,gp,bpz,bf,bog,zj}, and to
authors of more pedagogical articles \cite{bh,hs}. At the same time,
the confusion persists, and continues to crop up from time to time.
Thus it seems worthwhile to address this issue here.

It should also be stated at the outset that our article is of
no interest if one only wants the splitting
to ``exponential accuracy", i.e., if one only wants the Gamow factor.
In most physical applications, this is all that can
sensibly be done, and the prefactor is best estimated as an ``attempt
frequency." From a mathematical point of view, however, the tunnel
splitting is found as an asymptotic approximation in the limit
$\hbar\to 0$. An ``exponentially accurate" answer for the splitting
$\Dta$ is one for which one only has an asymptotically correct
result for $\ln\Dta$. An asymptotically correct answer for $\Dta$
itself requires worrying about the prefactor. More importantly, any
formalism which did not give this prefactor, or which was 
incapable of giving it correctly even in principle,
would be regarded as logically unsound. It is from this perspective
that the prefactor is important. The confusion that is referred to
above can be said to pertain solely to this prefactor, and readers
who are not concerned with such arcana should stop reading here.
On the other hand, the WKB analysis presented here is easily incorporated
into a graduate level discussion of tunnel splittings, with little
additional effort beyond that of the standard treatment,
and there is no reason not to do so. The problem may also serve
as a ``real-life" physics example of some nontrivial asymptotic analysis.
(To avoid misunderstanding, let us also state at the outset what we mean
by WKB theory. We use the term in the broad sense used
by Bender and Orszag \cite{bo}, or by Berry and Mount \cite{bm}, i.e.,
as a body of mathematical techniques that yield global
understanding and systematic asymptotic approximations to solutions of many
linear differential equations,
\begin{figure}
\centerline{\psfig{figure=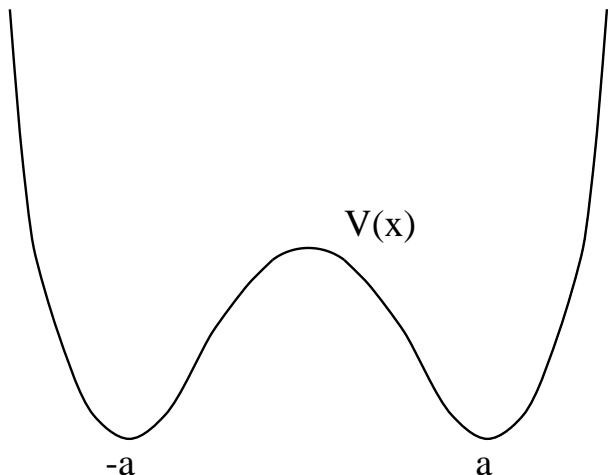}}
\caption{Symmetric double well potential $V(x)$ considered in this paper.
The minima at $x=\pm a$ are taken to be quadratic.}
\end{figure}
including Schr\"odinger's equation,
and shares ideas with other methods
such as asymptotic matching and patching, boundary layer theory, etc.
Some readers may, however, define WKB to encomapss only 
the approximate exponential form (\ref{psi01}) (and its oscillatory
counterpart) and connection formulas at linear turning points. 
A subset of these readers may recognize that our treatment in Sec.~II
is tantamount to the use of quadratic turning point \cite{qtp}
connection formulas (for which see Berry and Mount \cite{bm}),
and object that this goes beyond WKB.
Such a semantic restriction of the scope of the term `WKB', should in
our view, be avoided. Since the method did not originate with
Wentzel, Kramers, and Brillouin,
but had antecedents in the work of Jeffreys, Rayleigh, Carlini,
Green, Liouville, and perhaps others, there is no compelling
historical reason for this restriction (as opposed
to say, the usages `Einstein model' for lattice specific heat, or the
`Kronig-Penney model' for electrons in a periodic potential),
and it is surely more useful to present the
subject to students as one which allows for systematic improvement,
and is not a closed subject of study even today.)

It is with this motivation that we revisit the problem of calculating
tunnel splittings in symmetric double well potentials. The article
has three distinct aims. The first is to
carefully calculate the prefactor multiplying the exponential of
the turning-point-to-turning-point action integral in the standard
WKB expression for the splitting $\Dta$. The correct formula is
Eq.~(\ref{wkbok}) below.
Our second aim is to present another
formula [See Eqs.~(\ref{simp}--\ref{defA})] for the ground state
splitting \cite{fn2}, that reduces everything to the evaluation of two
integrals involving the potential $V(x)$, and does not require
looking up any asymptotic expansions. 
The last aim is to show
that the correct WKB and instanton \cite{jl,sc} methods do yield
the same ground state splitting. We will do this by starting with
Coleman's instanton method expression for the splitting \cite{sc}
and showing that it reduces to our simple formula.

We expect that our WKB based disucssion will be accessible to
students who have seen some graduate level quantum mechanics,
even though our starting formula for the splitting is rarely ever
mentioned in the common text books. One book where it does appear
is again that of Landau and Lifshitz \cite{ll}, who give an
exceptionally lucid and self-contained discussion. We will comment
further on this formula when we come to it in Sec. II, and give
a separate derivation of it in Appendix A. 
The instanton formalism, and the path integral approach to quantum
mechanics upon which it is in turn based, are less likely to be familiar,
but clear and accessible discussions have been given in this journal by
Holstein \cite{bh}, and by Holstein and Swift \cite{hs}.
The former discusses
exactly the same problem as us, namely, the tunnel splitting, but for
the special case of a quartic double well. We enthusiastically
recommend Coleman's somewhat longer but authoratitive article on
instantons \cite{sc} to readers who wish to learn more about this
technique. A related paper by Carlitz and Nicole \cite{cn} may also
be read profitably.

A plan of our paper and summary of our results is as follows.
We consider a potential which is
smooth, reflection symmetric about $x$=0 [$V(-x) = V(x)$],
and which has quadratic minima at $x=\pm a$. The correct WKB answer for the
ground state splitting is \cite{fn3}
\beq
\Dta = {\hbar\om \over \sqrt{e\pi}}
              \exp\biggl[ - \int_{-a'}^{a'} {|p|\over \hbar}dx
                         \biggr]. \label{wkbok}
\eeq
Here, $\om$ is the frequency of small amplitude oscillations in the
wells about $x = \pm a$, $\pm a'$ are the classical turning points
given by the equation
\beq
V(a') = E_0 = \hf\hbar\om  + o(\hbar), \label{ctp}
\eeq
and $p$ is the momentum
\beq
p(x) = [ 2m (V(x) - E) ]^{1/2}. \label{mom}
\eeq
Note that $p(x)$ is imaginary in the classically forbidden region
$ -a' < x < a'$, and that we have taken $V(\pm a) = 0$.

The proof of Eq.~(\ref{wkbok}) is given in Sec.~II. This entails
matching the WKB wavefunction in the classically
forbidden region to the exact harmonic oscillator wavefunction
near the classical turning point, rather than use the connection formulas.
This matching calculation may actually be found in a paper by
Furry~\cite{wf},
but we include it because it is very short, and so as to have the complete
argument in one place in a consistent notation.

Prefactor corrections similar to those in Eq.~(\ref{wkbok}) accompany
the excited states too, and are discussed in Appendix B.

The expression (\ref{wkbok}) is not easy to use because the 
integrand in the exponential is
close to a singularity near the limits, which means that the next
to leading dependence of $\Dta$ on $\hbar$ is not manifest. In fact, the
true preexponential factor varies as $\hbar^{1/2}$, and this is not
obvious from Eq.~(\ref{wkbok}). In Sec. III, we shall, therefore, carefully
extract the singular contributions to the action integral from the end
points, and show that the splitting may be written for a general potential
as
\beq
\Dta = 2\hbar \om \left( {m\om a^2 \over \pi\hbar} \right)^{1/2}
       e^A e^{-S_0/\hbar}, \label{simp}
\eeq
where $S_0$ (note the limits of integration) is the action integral
\beq
S_0 = \int_{-a}^a \left( 2mV(x) \right)^{1/2} dx, \label{S0}
\eeq
and
\beq
A = \int_0^a \biggl[ {m \om \over \sqrt{2mV(x)}} - {1 \over a-x}
               \biggr] dx . \label{defA}
\eeq
This expression does not have the complexities mentioned above. The
$\hbar^{1/2}$ dependence of the prefactor is apparent, and the
answer involves only integral functionals of $V(x)$.
These formulas will be applied to two model problems in Sec.~V.

In Sec. IV, we turn to the instanton approach \cite{jl,sc}.
Here, the splitting is expressed as
\beq
\Dta = 2\hbar K \left( {S_0 \over 2\pi\hbar} \right)^{1/2}
        \exp(-S_0/\hbar), \label{inst}
\eeq
where $S_0$ is as in Eq.~(\ref{S0}), and
\beq
K = \left[ {\det( -\ptl_\tau^2 + \om^2) \over
            \det'[-\ptl_\tau^2 + m^{-1} V''(x_{\rm cl}(\tau))]}
     \right]^{1/2}   \label{vvdet}
\eeq
is the ratio of fluctuation determinants. Here, $x_{\rm cl}(\tau)$
is the instanton, which obeys the (Euclidean) classical equation
of motion
\beq
m {d^2 x_{\rm cl}(\tau) \over d\tau^2} - V'(x_{\rm cl}) =0,
   \label{eom}
\eeq
with the boundary conditions $x_{\rm cl}(\pm\infty) = \pm a$,
and the additional condition $x_{\rm cl}(0) = 0$ to fix the
time translation degree of freedom. Further, $V' = dV/dx$,
$V'' = d^2V/dx^2$, and the prime on the det in Eq.~(\ref{vvdet})
means that the zero eigenvalue of the operator argument is to be
excluded from a product of all eigenvalues.

A short and pedagogical discussion of the instanton method
may be found in Holstein \cite{bh}. Holstein
does not explain how to calculate the quantity $K$ (which is
proportional to his quantity $K_1$), so a few remarks
that help elucidate its nature may not be out of place. Note first,
that each determinant appearing in Eq.~(\ref{vvdet}) is of a
one-dimensional Schrodinger-like operator, in which $\tau$
plays the role of position, and either $\om^2$ or
$m^{-1}V''(\xcl(\tau))$ plays the role of the potential energy.
Next we note, that
the determinant of such an operator may be defined as an infinite
product of all its eigenvalues. The spectrum of operators at hand
may be rendered completely discrete by adding hard walls at
$\tau = \pm T$, and letting $T\to\infty$ at the end. This scheme
has the advantage that it allows us to put the eigenvalues of both
operators in one-to-one correspondence. This in turn enables us
to argue that one eigenvalue is missing in the denominator, and since
the eigenvalues have the dimensions of $\ptl_{\tau}^2$ or $\om^2$,
we see that $K$ has the dimensions of frequency. For a smooth
potential $V(x)$, the curvature $m^{-1}V''(\xcl(\tau))$ is never very
different from $\om^2$ in the interval $-a \le x \le a$, so there is
only one frequency scale in the
problem. It follows that $K$ is of order $\om$.

To the author's knowledge, the equivalence of Eqs.~(\ref{wkbok})
and (\ref{inst}) has only been shown for particular examples, such
as the quartic double well potential. A general demonstration
is lacking, although it is implicit in the fact that both methods start
from the same point and are correctly executed. A direct demonstration
is therefore of some value. Secondly, Eq.~(\ref{inst}) is complicated
just as Eq.~(\ref{wkbok}) was. Indeed, the ratio of
determinants seems more intimidating and harder to calculate than the
integral in Eq.~(\ref{wkbok}). Langer's original calculation \cite{jl}
for $K$ is quite involved, and while it has been greatly simplified by
Coleman \cite{sc}, it still requires solving the auxilliary problem of
finding the instanton $x_{\rm cl}(\tau)$. 

To relate the instanton and WKB answers, ws therefore show in Sec. IV,
that Coleman's result for $K$ may be expressed
in terms of the integral $A$ in Eq.~(\ref{defA}). Thus, although we
do not provide a complete derivation of the instanton result for
$\Dta$, once Coleman's formulas are accepted, our analysis shows the
equivalence of the two approaches.

\section{WKB formula for ground state splitting}
\label{wkb}

Our starting point is the formula \cite{fn4}
\beq
\Dta = {2\hbar^2 \over m} \psi_0(0) \psi'_0(0), \label{herr}
\eeq
where $\psi_0(x)$ is an approximate solution to Schr\"odinger's equation
with energy $E_0$, which is localized in the right hand well, and
decays away from that well, in the {\it entire} central, classically
forbidden region. Further, it is normalized to yield unit total
probability in the right hand well. Note that it is not necessary to
examine or specify the behaviour of $\psi_0(x)$ near $x = -a'$.
Lastly, $\psi'_0(x) = d\psi_0(x)/dx$.

We digress briefly at this point to comment on the formula
(\ref{herr}), which does not seem to be widely known. It is sometimes
named after Conyers Herring, who derived it (in a slightly more general form,
in fact) in the course of evaluating
the {\it g\/}-{\it u} splitting of the two lowest electronic states of
the H$_2^+$ molecular ion \cite{ch,hf}. The actual derivation is simple,
and very similar to the usual half-page of analysis used to argue that the
the Hamiltonian operator in the position representation is Hermitean
and has real eigenvalues, and so we refer readers to Landau and
Lifshitz \cite{ll} for the details. Instead, we present an alternative
derivation in Appendix A.

Resuming our argument, we note that
the WKB approximation for $\psi_0(x)$
in the region $(a'-x) \gg (\hbar/m\om)^{1/2}$ is
\beq
\psi_0(x) = {C_0 \over \sqrt{|v(x)|}}
            \exp {1\over \hbar} \int_0^x | p(x')| dx', \label{psi01}
\eeq
where $v(x) = p(x)/m$, and $C_0$ is a constant to be found by matching
on to the solution in the well. To the accuracy of this solution,
$\psi'_0(x) \approx (m |v(x)|/\hbar) \psi_0(x)$, so that
\beq
\Dta = 2\hbar C_0^2. \label{Csq}
\eeq

To find $C_0$, we first note that near $x=a$, $\psi_0(x)$ is very
accurately given by the ground state harmonic oscillator wave function
\beq
\psi_0(x) = \left( {m\om \over \pi\hbar} \right)^{1/4}
             \exp\left[ -{m\om \over 2\hbar} (x-a)^2 \right].
                 \label{psi02}
\eeq
In fact, this form holds well into the forbidden region, and so
can be directly matched to Eq.~(\ref{psi01}) without invoking
connection formulas. In the overlap region, we may write
\beq
|p(x)| = m \om \left[ (a-x)^2 - u_0^2 \right]^{1/2},
                             \label{px}
\eeq
where
\beq
u_0 = a - a' \approx (\hbar/m\om)^{1/2}. \label{defu}
\eeq
The term $u_0^2$ in Eq.~(\ref{px}) may be neglected in evaluating the
$|v(x)|^{-1/2}$ prefactor in Eq.~(\ref{psi01}), so that we may
write
\beq
\psi_0(x) \approx {C_0 \over \sqrt{\om (a-x)}}
            \exp \biggl[ {1\over \hbar} \int_0^{a'} | p(x')| dx'
             + \Phi(x) \biggr] , \label{psi03}
\eeq
where
\bea
\Phi(x) &=& -{m \om \over \hbar}
              \int_x^{a'}
                \left[ (a-x')^2 - u_0^2 \right]^{1/2} dx' \nnu \\
    &\approx& - {m \om (a-x)^2 \over 2 \hbar}
              + \ln \left( {2(a-x) \over u_0} \right)^{1/2}
              + {1 \over 4} + O\left( {u_0 \over a-x} \right)^2.
\label{Phi}
\eea
Comparing Eqs.~(\ref{psi03}) and (\ref{Phi}) with Eq.~(\ref{psi02}),
we obtain
\beq
C_0 = \left( {\om^2 \over 4\pi e} \right)^{1/4} 
       \exp \biggl[ -{1\over \hbar}\int_0^{a'}|p(x)| dx \biggr].
   \label{ansC}
\eeq

Substituting Eq.~(\ref{ansC}) in Eq.~(\ref{Csq}), and making use of
the reflection symmetry of $V(x)$, we obtain
\beq
\Dta = {\hbar\om \over \sqrt{e\pi}}
              \exp\biggl[ - \int_{-a'}^{a'} {|p|\over \hbar}dx
                         \biggr], \label{wkbok2}
\eeq
which is Eq.~(\ref{wkbok}).

Landau and Lifshitz obtain $C_0$ by using connection formulas near a
linear turning point, and the standard WKB result for the normalization
of a bound state. For the ground state, this procedure is not
accurate enough, and we must proceed as above. (Some readers may
recognize in our procedure, the use of quadratic
turning point formulas \cite{bm}.) Once this is realized,
similar prefactor corrections are expected for the low lying excited
states. These corrections are obtained in Appendix B.

\section{Simpler Expression for Ground State Splitting}
\label{simpdel}

In this section we will show that the WKB expression
(\ref{wkbok}) leads to the simpler result
(\ref{simp}--\ref{defA}). The objective, clearly,
is to let the action integral run from $-a$ to $a$ instead of from
$-a'$ to $a'$. To that end, let us denote $y = a-x$, $V(x) = U(y)$,
and define, for a general $u$,
\beq
I(u) = \int_u^a \bigl[ 2m\bigl(U(y) - U(u)\bigr)\bigr]^{1/2} dy.
\label{defIu}
\eeq
The quantity $I(u_0)$ is clearly half the action integral in
Eq.~(\ref{wkbok}), i.e.,
\beq
I(u_0) = {1\over 2\hbar} \int_{-a'}^{a'} |p(x)| dx.
\label{actint}
\eeq
Our goal is to expand $I(u)$ for small $u$.
The analysis in Sec.~II reveals [see Eq.~(\ref{Phi})] that this
expansion contains a term varying as $u^2\ln u$, which means that
the expansion cannot be found by simply differentiating $I(u)$.
More precisely, the term of order $u^2 \ln u$ is easily found, but
the term of order $u^2$, which we also need, is rather harder to get.

We therefore resort to a subtraction and write
$I(u) = I_1(u) + I'_1(u)$, where
\bea
I_1(u)  &=& \int_u^a \sqrt{2m U(y)} dy, \label{I1def} \\
I'_1(u) &=& \int_u^a
            \biggl( \sqrt{2m[ U(y) - U(u)]} - \sqrt{2m U(y)}
               \,\biggr) dy. \label{I1pdef}
\eea
$I_1(u)$ can be directly expanded:
\bea
I_1(u)  &\approx& \int_0^a \sqrt{2m U(y)} dy 
                 -\int_0^u \left( m\om y + O(y^2) \right) dy \nnu \\
        &=& I(0) - \hf m \om u^2 + O(u^3), \label{I1ans}
\eea
while for $I'_1(u)$ we have, as $u \to 0$,
\beq
I'_1(u) \approx -(m\om u)^2 \int_u^a
         {dy \over  \sqrt{2m[ U(y) - U(u)]} + \sqrt{2m U(y)}}.
\label{I1p2}
\eeq
Note that in writing this expression, we have taken
$U(u) \approx m\om^2 u^2/2$ in the numerator, as $u$ is small.
It still can not be expanded directly, however, so we
perform another subtraction, and write $I'_1 = I_2 + I_3$,
where
\beq
I_2(u) = -\hf (m\om u)^2 \int_u^a
         {dy \over  \sqrt{2m[ U(y) - U(u)]}}.
\label{I2def}
\eeq
The difference $I_3$ contains a proportionality factor
$U(u)$, which we again approximate by $m \om^2 u^2/2$. This
leads to
\beq
I_3(u) \approx -\hf (m\om u)^4 \int_u^a
         {dy \over  \sqrt{2m[ U(y) - U(u)]} 
         \biggl[ \sqrt{2m[ U(y) - U(u)]} + \sqrt{2m U(y)}
                \biggr]^2 }.
\label{I3def}
\eeq

Let us consider $I_2(u)$ first. We can clearly write the integral
as
\beq
\int_u^a {dy \over m\om \sqrt{y^2 - u^2}}
+ \int_u^a \Biggl[
     {1\over \sqrt{2m[U(y) - U(u)]}}
             - {1\over m\om \sqrt{y^2 - u^2}}
           \Biggr] dy.
\label{I22}
\eeq
The first integral can be done exactly, while in the second we can
set $u=0$ inside the integrand and in the limits to leading order. We thus find
\beq
I_2(u) \approx -\hf m \om u^2 \ln{2a \over u}
        -\hf (m \om u)^2 \int_0^a
          \left[ {1\over \sqrt{2mU(y)}} - {1\over m\om y}
          \right] dy
        + O(u^4\ln u).
\label{I2ans}
\eeq
      
The leading behaviour of $I_3(u)$, on the other hand, is controlled
by the lower limit in the integral (\ref{I3def}), where we can
again approximate $U(y)$ by $m \om^2 y^2/2$. Hence,
\bea
I_3(u) &\approx& \hf m\om u^4
         \int_u^a {dy \over \sqrt{y^2 - u^2}
                       \left[ \sqrt{y^2-u^2} + y \right]^2} \nnu \\
       &\approx& {1\over 4} m \om u^2 + O(u^4).
\label{I3ans}
\eea

Collecting together Eqs.~(\ref{I1ans}), (\ref{I2ans}), and (\ref{I3ans}),
and putting $u = u_0 = (\hbar/ m\om)^{1/2}$, we get
\beq
I(u_0) \approx I(0) - {\hbar \over 2} \ln{2a \over u_0}
           - {\hbar \over 4}
           -{\hbar \over 2} \int_0^a
                 \Biggl[ {m\om \over \sqrt{2mU(y)}} - {1\over y}
                 \Biggr] dy + \cdots.
\label{Iu0ans}
\eeq
The last integral in the above equation is nothing but the quantity
$A$ defined in Eq.~(\ref{defA}). Using $u_0 = (\hbar/m\om)^{1/2}$
once again, and putting together Eqs.~(\ref{Iu0ans}), (\ref{actint}),
and (\ref{wkbok}), we get
\beq
\Dta = {2\hbar\om \over \sqrt{e\pi}}
         \left( {m \om a^2 \over \hbar} \right)^{1/2}
       \sqrt{e} e^A e^{-2I(0)/\hbar}.
\label{Dtaans2}
\eeq
Since $I(0) = S_0/2$, this is nothing but Eq.~(\ref{simp}).

\section{Equivalence of WKB and Instanton Results}
\label{equival}

Our next step is to show that the instanton result (\ref{inst})
also leads to Eqs.~(\ref{simp}--\ref{defA}),
and thus prove the equivalence of the WKB and instanton results for the
ground state splitting. To do this, we use Coleman's
result for the ratio $K$ \cite{sc}. According to him,
\beq
K = \sqrt{2\om} \be, \label{Kbe}
\eeq
where $\be$ is related to the asymptotic, $\tau\to\pm\infty$
behaviour of the instanton velocity via
\beq
x_1(\tau) \equiv \left( {m\over S_0} \right)^{1/2}
                 {d x_{\rm cl} \over d\tau}
          \approx \be e^{\mp \om\tau}
           \qquad\hbox{as $\tau\to\pm\infty$.}
\label{x1be}
\eeq

It is easy to integrate the equation of motion (\ref{eom}) for the
instanton, and obtain $x_1(\tau)$. Using the fact that
$\xcl(0) = 0$, we have
\beq
\tau = m \int_0^{\xcl} {dx \over \sqrt{2m V(x)}}.
\label{tvsx}
\eeq
This diverges as $\xcl \to a$, as it should. We extract the
divergence by subtracting and adding the leading singular part
of the integrand. This yields,
\bea
\tau &=& m\int_0^{\xcl} \left[ {1\over \sqrt{2m V(x)}}
                             - {1 \over m \om (a-x)}
                      \right] dx
        + {1\over\om} \ln\left( {a \over a-\xcl} \right) \nnu \\
     &\approx& 
        {1\over\om} \ln\left( {a \over a-\xcl} \right) + {A\over \om}
           \qquad\hbox{as $\xcl\to a$.} \label{tvsxans}
\eea
$A$ is, of course, the quantity defined in Eq.~(\ref{defA}).

Thus, as $\tau\to \infty$,
\bea
\xcl(\tau) &\approx& a - a e^A e^{-\om\tau},  \label{xclas} \\
{d\xcl \over d\tau} &\approx& a\om e^A e^{-\om\tau}.
\label{dxclas}
\eea
Comparing with Eq.~(\ref{x1be}), we can read off $\be$ immediately:
\beq
\be = a \om \left( {m \over S_0} \right)^{1/2} e^A.
\label{beans}
\eeq
Hence,
\beq
K = a \om \left( {2m \om \over S_0} \right)^{1/2} e^A,
\label{Kans}
\eeq
and
\beq
\Dta = 2 \hbar\om \left( {m\om a^2 \over \pi \hbar} \right)^{1/2}
                  e^A e^{-S_0/\hbar},
\label{Dta3}
\eeq
which is what we set out to show.

\section{Illustrative Examples}
\label{examp}

We conclude with two elementary examples to which we apply
Eqs.~(\ref{simp}--\ref{defA}).

The first example is that of the quartic double well,
\beq
V(x) = V_0(x^2 - a^2)^2/a^4. \label{Vquar}
\eeq
Here, $V_0$ is the barrier height. The frequency $\om$ is given by
\beq
\om = (8V_0/ma^2)^{1/2}. \label{omquart}
\eeq
It is simple to perform the integrals, and show that
\bea
{S_0 \over \hbar} &=& {4\over3} \left(
                                 {2mV_0a^2 \over \hbar^2}\right)^{1/2}
                      = {16\over 3}{V_0\over\hbar\om}, \label{S0quar} \\
A &=& \ln 2. \label{Aquar}
\eea
Further, $(m\om a^2 /\pi\hbar)^{1/2} = (3S_0/2\pi\hbar)^{1/2}$, so
that
\beq
\Dta = 4\sqrt3 \hbar\om \left( {S_0\over 2\pi\hbar} \right)^{1/2}
                          e^{-S_0/\hbar}, \label{splquar}
\eeq
which is a well known form for the answer.

Our second example involves the spin Hamiltonian
\beq
\ham = -\gam S_z^2 - \al S_x, \label{Hspin}
\eeq
where $S_\al$ ($\al = x, y, z$) are components of the dimensionless
spin operator $\bS$ obeying the commutation rules
$[S_\al, S_\be] = i\eps_{\al\be\gam} S_\gam$, and we also take
$\al > 0$, $\gam > 0$. 

In the limit where the magnitude $S$ of the
spin is very large \cite{fn5}, Eq.~(\ref{Hspin}) can be viewed as tending
to a classical Hamiltonian,
wherein the dynamics are defined by giving the Poisson brackets
$[S_\al,S_\be]_{\rm PB} = \eps_{\al\be\gam}S_\gam$.
The classical energy has two degenerate minima when the spin lies in the
{\it xz} plane at angles $\tta_0$ or $\pi - \tta_0$ to the {\it z} axis,
where
\beq
\sin\tta_0 = \al/2\gam S. \label{tta0}
\eeq
For the quantum mechanical problem, with large but finite $S$, we expect
the ground states around these classical orientations to be admixed by
tunneling, thus giving rise to a small splitting of the energy levels.

Compared to the tunneling of massive particles with a position
coordinate, the tunneling of spins is somewhat less familiar, but it is
a perfectly bona fide instance of the general tunneling phenomenon.
Spin tunnel splittings have been calculated by a variety
of means for some time now \cite{es,vs}, but obtaining the
prefactors in the tunnel splittings correctly to order $S^0$ as
$S\to\infty$, has proven to be a fairly difficult task, and the
instanton calculations \cite{es,bpp} are especially subtle. The
discrete WKB method uses only elementary methods of analysis, but 
the calculations to date \cite{vs,ag} are still lengthy. There is an
even simpler method, however, which is due to Scharf, Wreszinski, and van
Hemmen \cite{SWv}. These authors expand a general state $\ket\psi$
in the $S_z$ eigenbasis $\{\ket m\}$ ($S_z\ket m = m\ket m$) as
$\ket\psi = \sum_m D_m \ket m$, and construct a generating function
$\sum_m D_m x^m$. After a few changes of variables, the
Schr\"odinger equation for $\ket\psi$ is turned into the following
Schr\"odinger equation [See their Eq.~(2.17)] for a wavefunction $y(z)$
related to the generating function:
\beq
-\gam^2 {d^2 y \over dz^2} + V(z) y(z) = \gam E y(z), \label{Seq}
\eeq
with the potential
\bea
V(z) &=& {1\over 4}\al^2(\cosh z - \cosh z_0)^2, \label{Vz} \\
\cosh z_0 &=& (2S + 1)\gam/\al. \label{z0}
\eea
The quantity $E$ in Eq.~(\ref{Seq}) is the energy eigenvalue of the
Hamiltonian (\ref{Hspin}). 

The potential (\ref{Vz}) is even about $z=0$, and has minima at $\pm z_0$.
Our formalism is directly
applicable if we identify $\hbar^2/2m$ with $\gam^2$, and $m\om^2$
with $V''(\pm z_0) = \hf \al^2 \sinh^2 z_0$. The action integral
(\ref{S0}) is easily seen to be
\bea
{S_0\over\hbar} &=& {\al\over\gam}
                       \int_0^{z_0}(\cosh z_0 - \cosh z) dz \nnu \\
                &=& {\al\over\gam} (z_0\cosh z_0 - \sinh z_0),
                       \label{S0spin}
\eea
while the correction (\ref{defA}) is given by
\beq
A = \int_0^{z_0} \left[
        {\sinh z_0 \over \cosh z_0 - \cosh z}
           - {1 \over z_0 - z} \right] dz. \label{Aspin}
\eeq
Since the integrand is nonsingular at $z=z_0$ (or if one so wishes
to view matters, has a removable singularity at that point),
the integral can be found by changing the upper limit to
$z_0 - \delta$, integrating the two terms separately, and letting
$\delta \to 0$ at the end. The integrals involved are elementary, and
the final result is
\beq
A = \ln \left( {2 \sinh z_0 \over z_0} \right). \label{Aspans}
\eeq

It remains to substitute Eqs.~(\ref{S0spin}) and (\ref{Aspans})
into our general formula (\ref{simp}). Recalling the
factor of $\gam$ on the right hand side of Eq.~(\ref{Seq}),
and the equivalences $\hbar^2/2m \equiv \gam^2$,
$m\om^2 \equiv V''(z_0)$, we obtain
\beq
\Dta = {4 \al^{3/2} \sinh^{5/2}z_0 \over \sqrt{2\pi\gam}}
         \exp \left[ -{\al\over\gam}
                     (z_0 \cosh z_0 - \sinh z_0) \right].
       \label{Dtaspin}
\eeq
It is easier to interpret this result if it is cast in terms of
the angle $\tta_0$ in Eq.~(\ref{tta0}). After a certain amount of
tedious algebra, one obtains, correct to order $S^0$ as $S\to\infty$,
\beq
\Dta = {8 \gam S^{3/2}\cos^{5/2}\tta_0 \over \pi^{1/2} \sin\tta_0}
      \lp {1-\cos\tta_0 \over 1+ \cos\tta_0} \rp^{S + 1/2} e^{2S\cos\tta_0}.
\label{split2}
\eeq
This is precisely the form quoted in Ref.~\cite{ag}.
Its relation to previous calculations is discussed there.

From the viewpoint of this article, the interesting comparison is
with Scharf, Wreszinski, and van Hemmen \cite{SWv}. These authors only
consider the case where $\al$ and $\gam$ are fixed as $S\to\infty$, so
that $\tta_0 = O(1/S)$. In that limit, they follow the Landau-Lifshitz
prescription, and obtain a ground state splitting
$\Dta' = (2S/\pi)e^J$, with
\beq
J = -2S\ln \left({8\over \al}\gam^2 S^2 \right) + 2S
     + (2S + 1) \ln \left[ \gam\bigl(2S + \tshf\bigr) \right]
     -\hf \ln\left[ \gam^2 \bigl(S + {\tst{1\over 4}}\bigr) \right].
\label{SWvHans}
\eeq
With a little work one can show that $\Dta'/\Dta$ is exactly
$(e/\pi)^{1/2}$. The discrepancy can in fact be seen in the
comparison between numerical and analytical results in Table 1
of Ref.~\cite{SWv}.
While $(e/\pi)^{1/2} = 0.930$, the ratio of their analytical answer for the
splitting to the numerical one (see the last two columns) decreases
from 0.987 to 0.951 as $S$ increases from 5 to 11, which is quite close.
A Richardson transformation \cite{fn6} of this ratio does suggest that it
tending to 0.93, but one cannot be certain of this conclusion, as there are
two erratic values at 0.90 and 0.91.

\acknowledgments
This work is supported by the National Science Foundation via grant
number DMR-9616749.

\appendix
\section{Herring's formula}

Let us denote the two states localized in the left and right wells by
$\ket R$ and $\ket L$ respectively. These states are degenerate in the
absence of tunneling with an energy $E_0$. The tunnel splitting is $\Dta$.
The Hamiltonian matrix in this two-state subspace is given by
\beq
\ham = \left(\matrix{ E_0 & \Dta/2 \cr
             \Dta/2       & E_0 \cr}\right)
      \label{ham22}
\eeq
Assuming that the system starts in the state $\ket R$ at $t=0$, it is
straightforward to show that the probability $P_R(t)$ for finding it in
the same state at a later time $t$ is given by
\beq
P_R(t) = \cos^2(\Dta t/2\hbar). \label{PRt}
\eeq
In particular,
\beq
{dP_R \over dt} = -{\Dta \over 2\hbar}
                   \sin\left({\Dta t \over \hbar}\right).
         \label{dPRt}
\eeq

To relate this abstract space description to that in position space,
we make use of the continuity equation for
probability. For a general wave function 
$\psi(x,t)$ that obeys Schr\"odinger's equation,
the probability density $P(x,t) = |\psi(x,t)|^2$ obeys 
\beq
{\ptl P(x,t) \over \ptl t} = - {\hbar \over m} {\ptl\over \ptl x}
                       {\rm Im}[\psi^*(x,t) \psi'(x,t)],
   \label{Pcont}
\eeq
where $\psi' \equiv \ptl\psi/\ptl x$. The probability $P_R(t)$ for being in the
right well is then given by
\beq
P_R(t) = \int_0^{\infty} P(x,t) dx. \label{PRt2}
\eeq
Differentiating with respect to $t$, making use of the continuity equation,
and the fact that $\psi(x,t) \to 0$ as $x\to\infty$ for any well behaved
solution, we obtain
\beq
{dP_R(t) \over dt} = {-\hbar \over m} {\rm Im}[\psi^*(0,t) \psi'(0,t)].
         \label{dPRt2}
\eeq

Let us finally consider the states $\ket R$ and $\ket L$
in position space. It is evident that
we should take $\tran xR$ to be the function described as $\psi_0(x)$
in Sec. II. By symmetry, $\tran xL = \psi_0(-x)$, and the energy eigenstates
are $(\psi_0(x) \pm \psi_0(-x))/\sqrt2$. It then follows that with the same
initial conditions as above, the wave function at an arbitrary time $t$ is
given by
\beq
\psi(x,t) = \psi_0(x)\cos(\Dta t/2\hbar)
             + i \psi_0(-x) \sin(\Dta t/2 \hbar). \label{psixt}
\eeq
Hence,
\beq
{\rm Im}[\psi^*(0,t) \psi'(0,t)] = \psi_0(0) \psi'_0(0) \sin(\Dta t/\hbar).
   \label{current}
\eeq
Substituting this result in Eq.~(\ref{dPRt2}) and comparing with
Eq.~(\ref{dPRt}), we obtain
\beq
\Dta = (2\hbar^2/m) \psi_0(0) \psi'_0(0), \label{herr2}\
\eeq
which is Eq.~(\ref{herr}), Herring's formula.

It is apparent from our derivation that Herring's formula holds whenever
the energy eigenfunctions are well approximated by the combinations
$(\psi_0(x) \pm \psi_0(-x))/\sqrt2$. (More precise statements of the
conditions for its validity are given in Ref.~\cite{ch}.)
Use of this formula greatly simplifies
the labour required to solve all the standard double-well problems: the double
square well with infinite side walls \cite{ago,gb},
the double delta function \cite{gb}, and
the double harmonic oscillator with a kink \cite{em}. For smooth potentials,
where WKB is indicated, it is far superior to the approach where one uses
connection formulas at all four turning points \cite{dp}. Further, the
formula brings out the physical point that a tunnel {\it splitting} is
intimately related to a tunneling {\it amplitude}, since it relates the
amplitude to make a transition between wells per unit time, $-i\Dta/\hbar$,
to the probability current.

The reader will also have noticed that the above argument cannot be carried
out for an asymmetric
potential, and indeed the very concept of tunnel splitting is then very delicate.
A tunneling amplitude can of course still be defined, but since this
amplitude is exponentially small in general (on account of the Gamow
factor), mixing between left and right well states will be negligible
unless the bottoms of the two wells are tuned to the same exponential
sensitivity. The mathematical formulation of these points leads to extremely
unpleasant transcendental equations, and the situation is not significantly improved 
in the case of symmetric potentials
if one approaches the problem solely in terms of enforcing continuity of the wave
function and its derivative. It is probably because of this fact that most
introductory or intermediate quantum mechanics texts
do not consider double-well tunnel splittings when they
discuss the WKB method. Use of Herring's formula (with or without
WKB) would make the problem much more tractable, and its widespread adoption is
thus greatly to be desired.

\section{WKB formula for higher state splittings}

It is to be expected that the formula given by Landau and
Lifshitz \cite{ll} is increasingly accurate for higher pairs
of states, assuming of course, that the WKB approximation is still
applicable. We will find that this is indeed so.
The splitting for the $n$th pair of levels, $\Dta_n$, is given by
\beq
\Dta_n = g_n {\hbar\om \over \pi}
              \exp\left[ - \int_{-a'_n}^{a'_n} {|p|\over \hbar}dx
                         \right], \label{Dtan}
\eeq
where $\pm a'_n$ are the classical turning points for the $n$th
energy level pair $(n+\hf)\hbar\om$, and
\beq
g_n = {\sqrt{2\pi} \over n!} \lp n + \tshf \rp^{n +\hf}
         e^{-(n + \hf)}. \label{gndef}
\eeq
Note that if $g_n$ were unity, we would have the formula of
Ref.~\cite{ll}. The corrections are indeed small:
$g_0 = (\pi/e)^{1/2} \approx 1.075$, $g_1 \approx 1.028$,
$g_2 \approx 1.017$, and so on. Stirling's formula for $n!$
shows that $g_n \to 1$ as $n \to \infty$.

The derivation of Eq.~(\ref{Dtan}) and the conditions under
which it holds are straightforward though long. The starting
point is still Herring's formula, Eq.~(\ref{herr}). The
procedure of Sec.~II can be applied word for word, with $a'$,
$u_0$, and $C_0$ replaced by $a'_n$, $u_n$, and $C_n$,
respectively. The essential part 
is to find $\psi(x)$ in the classically
forbidden region and match it onto the $n$th harmonic oscillator
state in the vicinity of the well $x \approx a$. We leave it
as an exercise --- or see Ref.~\cite{wf} ---
to show that the function $\Phi(x)$, defined
in Eq.~(\ref{psi03}), is given in the overlap region by
\beq
\Phi(x) = -\hf {m\om \over \hbar} (a-x)^2
          + {2n + 1 \over 2} \ln\lp {2(a-x) \over u_n}\rp
          + {2n + 1 \over 4} + \cdots, \label{Phin}
\eeq
so that the leading approximation to $\psi_0(x)$ is given by
$C'_n(a-x)^n \exp(-x^2/2u_0)$, where
\beq
C'_n = {C_n \over \om^{1/2}} \lp 2\over u_n\rp^{n + \hf}
          \exp{\lp {1\over \hbar} \int_0^{a'_n} |p| dx
                + {2n + 1\over 4} \rp} \label{Cprn}
\eeq
This is of the precise form required to be match on to the
$n$th excited state harmonic oscillator wave function:
\beq
\psi_0(x) \approx \lp {m \om \over \pi\hbar} \rp^{1/4}
             {2^{n/2} \over \sqrt{n!}} e^{-\xi^2/2}
             \lp \xi^n + O(\xi)^{n-2} \rp,
     \label{psin}
\eeq
with $\xi \equiv (a-x)/u_0$. The match yields the constant $C_n$,
and the splitting $\Dta_n$, which equals $2\hbar |C_n|^2$, is easily
shown to be given by Eq.~(\ref{Dtan}) and (\ref{gndef}).

The calculation sketched above assumes that the turning point
and the wavefunction in its vicinity are well approximated by
taking the potential well to be parabolic. This is clearly less
accurate as $n$ gets large. Defining $y$ and $U(y)$ as
in Sec.~IV, let us keep the cubic and quartic terms in $y$ in
$U(y)$:
\beq
U(y) = \hf m \om^2 y^2 + \al y^3 + \be y^4.
\label{U4}
\eeq
The wavefunction for the $n$th state can be found by perturbation
theory assuming that $\al$ and $\be$ are small. The key requirement
is that the correction terms be small compared to the unperturbed
wavefunction near the turning point $y = u_n$. The dominant
corrections are those that entail the Hermite polynomials of order
$n+1$ through $n+4$. Since the coefficient of the largest
power of $x$ in these polynomials is known, the conditions for the
corrections to be small are not difficult to find. They are
\bea
\al &\ll& {6 \over 13n + 11} (2n + 1)^{-1/2}
              \lp {m^3 \om^5 \over \hbar}\rp^{1/2}, \label{conda} \\
\be &\ll& {4 \over (6n+7)(2n+1)} {m^2 \om^3 \over \hbar}.
          \label{condb}
\eea
It is also possible to show that when these conditions hold, the
corrections to $u_n$ and the energies of the harmonic oscillator
states are negligible, so that the conditions are self-consistently
derived.

It need hardly be said that the conditions (\ref{conda}) and
(\ref{condb}) are only necessary, not sufficient,
for Eq.~(\ref{Dtan}) to apply. In addition, one must have the
requirement for the WKB approximation itself to hold, which in
the present context can be stated as $(n + \hf)\hbar\om
\ll V(0) - V(a)$.


\begin{references}
\bibitem{fn1} Indeed, some of the confusion about the
WKB results is perhaps due to the great authority of Ref. \cite{ll}.
For example,
many workers in the field of macroscopic quantum tunneling,
including the author, have often been puzzled and frustrated
that the final formula in this book does not yield the same ground
state splitting as the instanton approach for the quartic double well.
An erudite colleague who
studies chaos assisted tunneling has stated to the author that ``WKB is
known to be wrong for the ground state splitting."
\bibitem{qtp} We should note that the complete discussion of
quadratic turning point formulas is far more complex than needed for
our purpose. For us, the discussion in Sec.~II is sufficient, and only
draws upon knowledge of harmonic oscillator eigenfunctions.
\bibitem{fn2} Although we have not seen this formula elsewhere
it seems very likely that the sophisticated mathematical physicists who have
studied the relationship between WKB methods, instantons, and
large order perturbation theory \cite{sc,bw,gp,bpz,bf,bog,zj} would find it
familiar or obvious. Hence,
we hereby explicitly disavow any claims to originality.
\bibitem{fn4} This result
appears two displayed equations above the final result in Ref.~\cite{ll}.
As discussed there, and as may also be seen from our Appendix B,
it also holds for the splitting of any higher pair
of levels, if $E_0$ is replaced by the mean energy of the pair.
\bibitem{fn5} $S$ is defined in the usual way, i.e., such that the
eigenvalue of $\bS\cdot\bS$ is $S(S+1)$.
\bibitem{fn6} If the ratio for spin $S$ is denoted by $r_S$, the transformed
sequence is ${\tilde r}_S = (S+1)r_{S+1} - Sr_S$.
If $r_S \simeq r_{\infty} + O(1/S)$, then
${\tilde r}_S \simeq r_{\infty} + O(1/S^2)$.
\end{references}
\end{document}